\documentclass[12pt]{article}
\usepackage{a4wide}
\usepackage{amssymb,amsmath}
\usepackage{cite}
\usepackage[hyperindex=true,
          pdfstartview=FitH,
          bookmarksnumbered=true,
          bookmarksopen=true,
          citecolor=blue,
          linkcolor=blue,
          colorlinks=true,
          pdfborder=001,
          unicode]{hyperref}

\usepackage{booktabs}
\usepackage[font=small,labelfont=bf]{caption}

\usepackage{makecell}
\setcellgapes{0.5ex}
\newcommand\disy{\displaystyle}

\allowdisplaybreaks
\numberwithin{equation}{section}

\begin{document}
\title{A Note on Entropy Relations of Black Hole Horizons}
\author{Xin-he Meng$^{1,2}$\thanks{{\em
        email}: \href{mailto:xhm@nankai.edu.cn}{xhm@nankai.edu.cn}}\ ,
        Wei Xu$^1$\thanks{{\em
        email}: \href{mailto:xuweifuture@mail.nankai.edu.cn}{xuweifuture@mail.nankai.edu.cn}}\ , and
        Jia Wang$^1$\thanks{{\em
        email}: \href{mailto:wangjia2010@mail.nankai.edu.cn}{wangjia2010@mail.nankai.edu.cn}}\\
$^1$School of Physics, Nankai University, Tianjin 300071, China \\
$^2$State Key Laboratory of Theoretical Physics, ITP-CAS, \\
Beijing 100190, P. R. China}
\date{}
\maketitle
\begin{abstract}
We focus on the entropy relations of black holes in three, four and higher dimensions. These entropy relations include entropy product, ``part'' entropy product and entropy sum. We also discuss their differences and similarities, in order to make a further study on understanding the origin of black hole entropy at the microscopic level.
\end{abstract}

\section{Introduction}
One of the major challenges in quantum theories of gravity in the past years is understanding the origin of black hole entropy at the microscopic level, which is also a clue for probing the microscopics of black holes. For this aim, much attention had been paid to the additional entropy relations of black holes in thermodynamics. These entropy relations include entropy product \cite{Cvetic:2010mn,Toldo:2012ec,Cvetic:2013eda,Lu:2013ura,Chow:2013tia,
Detournay:2012ug,Castro:2012av,Visser:2012zi,Chen:2012mh,Castro:2013kea,Visser:2012wu,Abdolrahimi:2013cza,Pradhan:2013hqa,
Castro:2013pqa,Faraoni:2012je,Lu:2013eoa,Anacleto:2013esa}, ``part" entropy product \cite{Visser:2012wu,Wang:2013nvz} (i.e. $\sum_{1\leq i<j\leq D}(S_{i}S_{j})^{\frac{1}{d-2}}$, where $D$ and $d$ are the number of horizons and the dimensions respectively.) and entropy sum \cite{Wang:2013smb,Xu:2013zpa}, which are expected to not only be expressed solely in terms of the quantized charges including the electric charge $Q$, the angular momentum $J$, and the cosmological constant $\Lambda$ (which can be treated as pressure after explaining the
mass of the black hole as enthalpy rather than internal energy of the system), but also have the mass independence. These entropy relations are always introduced in black hole spacetime with multi-horizons including the physical horizons: the event horizon, inner (Cauchy) horizon and cosmological horizon; and the un-physical ``virtual'' horizons.

Consider the thermodynamics of other horizons and the physics inside the black hole is
not just an artificial game to play with. There exist several reasons why people study the thermodynamics of all horizons. Firstly, it is found that the Green functions are sensitive to the geometry near all the black hole horizons, and not just the outermost one \cite{Cvetic:1997uw,Cvetic:1997xv,Cvetic:2009jn}. The thermodynamic properties, especially for the entropy at each horizon, can therefore be expected to play a role in governing the properties of the black hole at the microscopic level. Besides, the parallel study, the entropy inequalities of multi-horizons of four dimensional general axisymmetric stationary solutions in Einstein-Maxwell theory \cite{Ansorg:2010ru,Hennig:2009aa,Ansorg:2009yi,Hennig:2008zy,Ansorg:2008bv,Ansorg:2007fh,Jaramillo:2007mi} are interpreted as a general criterion for extremality \cite{Booth:2007wu}. It also results in a No-Go theorem for the possibility of force balance between two rotating black holes \cite{Hennig:2010hz}. This makes the physics at each horizon more intriguing. On the other hand, in the study on the entropy relations, one need to include the necessary effect of the un-physical ``virtual" horizons, in order to preserve the mass independence \cite{Visser:2012wu,Wang:2013smb,Xu:2013zpa}. Only in this way, these additional equalities of multi-horizons of black holes are ``universal''. Furthermore, by using of these thermodynamical relations, the construction of thermodynamics for inner horizon of black hole catches more attentions \cite{Detournay:2012ug,Castro:2012av,Chen:2012mh,Pradhan:2013hqa,Castro:2013pqa,Pradhan:2013xha,Ansorg:2008bv,Ansorg:2009yi,Ansorg:2010ru}, which makes the properties of inner of black hole more interesting.

The study on the entropy product of multi-horizons black hole is generalize to many theories, including the super-gravity model \cite{Cvetic:2010mn,Toldo:2012ec,Cvetic:2013eda,Lu:2013ura,Chow:2013tia}, Einstein gravity \cite{Detournay:2012ug,Castro:2012av,Visser:2012zi,Chen:2012mh,Castro:2013kea,Visser:2012wu,Abdolrahimi:2013cza,Pradhan:2013hqa} and other modified gravity models \cite{Castro:2013pqa,Cvetic:2013eda,Faraoni:2012je,Lu:2013eoa,Anacleto:2013esa} in both four and higher dimensions. It is always independent of the mass of the black hole \cite{Cvetic:2010mn,Castro:2012av,Toldo:2012ec,Chen:2012mh,Visser:2012zi,Cvetic:2013eda,Abdolrahimi:2013cza,Lu:2013ura,
Anacleto:2013esa,Chow:2013tia,Castro:2013kea,Lu:2013eoa,Wang:2013smb,Xu:2013zpa}. However, the mass independence of entropy product fails in some asymptotical non-flat spacetime \cite{Faraoni:2012je,Castro:2013pqa,Detournay:2012ug,Visser:2012wu}. Hence, the ``part'' entropy produce and entropy sum are introduced, which always are independent of the mass of the black hole in (A)dS spacetime \cite{Visser:2012wu,Wang:2013nvz,Wang:2013smb,Xu:2013zpa} (and seem to be more ``universal''). To say more accurately, the latter two entropy relations never depend on the electric charge $Q$ and angular momentum $J$, but the cosmological constant and the constants characterizing the strength of these extra matter field.

In this paper, we firstly present all entropy relations including entropy product, ``part'' entropy product and entropy sum in three dimensions, which is never studied in previous literature, in order to improve the study on entropy relations. Then we revisit some known entropy relations and give some new unknown ones in four and higher dimensions. We discuss their differences and similarities in general dimensions, in order to make a further study on understanding the origin of black hole entropy at the microscopic level. After having a holistic look at the entropy relations, we conclude that in the theory without Maxwell field entropy product and ``part'' entropy product belong to the same kind of entropy relation, because the mass independence of entropy product and ``part'' entropy product hold complementary. In rotating (A)dS spacetime, entropy product is independent of mass, while reducing to static (A)dS spacetime ``part'' entropy product takes the mass independence. When the case in static (A)dS spacetime reduces to flat spacetime, electric charge $Q$ plays a switch role to the dependent of mass or not of entropy product, while ``part'' entropy product turns to depend on mass \cite{Wang:2013nvz}. However, there are two kind of failed examples in which the mass independence of ``part'' entropy product or entropy product disappears: the degenerated cases with only two horizons, i.e. the cases that entropy product and ``part'' entropy product merge into the same entropy product and have some vanishing charges (electric charge and angular momentum, and cosmological constant) meanwhile, such as the static uncharged BTZ black hole and Kerr-Newman black hole in Gauss-Bonnet theory \cite{Castro:2013pqa}; since they do not admit the entropy area law. For the entropy sum, it only depends on the constant characterizing the strength of the background spacetime (the cosmological constant etc.) and never depends on electric charge, angular momentum and mass. Consider the ``part" entropy product and entropy sum together, we find that they both have solely the constant (characterizing the strength of the background spacetime, such as the cosmological constant etc.) dependence, other than the electric charge and angular momentum dependence, which is different from the entropy product. It is found that ``part'' entropy product and entropy sum of Schwarzschild-de-Sitter black hole are actually equal, when only the effect of the physical horizons are considered, as they both can be simplified into a mass independent entropy relations of physical horizon. This also reveal that one can explaining the origin of black hole entropy at the microscopic level without considering the effect of the un-physical virtual horizons. For the sake of brevity, this idea is only checked in Schwarzschild-de-Sitter black hole, while the calculations for the other cases (the Kerr-Newman-(Anti-)de-Sitter black holes etc.) can be performed in a similar manner.

This paper is organized as follows. In the next Section, we will investigate the entropy relations
of $2+1$ dimensional black holes. In Section 3, we take as a holistic look at the entropy relations in four and higher dimensions and discuss their differences and similarity in general dimensions. Section 4 is devoted to the conclusions and discussions.

\section{Entropy relations of black holes with multi-horizons in $(2+1)$ dimensions}
Entropy relations including entropy product, ``part'' entropy product and entropy sum in four and higher dimensions are studied widely. In this section, we present all entropy relations of $(2+1)$ dimensional BTZ black hole \cite{Banados:1992wn} and hairy black hole \cite{Martinez:1996p2505,Henneaux:2002wm,Nadalini:2007p2561,Xu:2013nia}, which is never studied in previous literature, in order to improve the study on entropy relations. These black holes are special and different from that in four and high dimensions, because in $(2+1)$ dimensions, smooth black hole horizons can exist only in the presence of a negative cosmological constant \cite{Ida}.

\subsection{Entropy relations of BTZ black hole}
Consider the BTZ black hole \cite{Banados:1992wn}, which is the solution of Einstein equations in the theory with the lagrangian
\begin{align}
\mathcal{L}=\frac{1}{2\pi}\int d^3x\sqrt{-g}\left(R-2\Lambda\right),
\end{align}
where the cosmological constant $\Lambda=-\frac{1}{l^2}$. The metric takes the form as
\begin{align*}
ds^2=-f(r)dt^2+\frac{dr^2}{f(r)}+r^2(N^{\phi}(r)dt+d\phi)^2,
\end{align*}
with the horizon function $f(r)$ and the angular velocity $N^{\phi}(r)$
\begin{align*}
f(r)&=-M+\frac{r^2}{\ell^2}+\frac{J^2}{4r^2}, \\
N^{\phi}(r)&=-\frac{J}{2r^2},
\end{align*}
where $M$ and $J$ are the mass and angular momentum of the black hole respectively.
We are interested in the entropy relations of the multi-horizons, which are the roots of $f(r)$ and read as
\begin{align*}
r_1&=\sqrt{\left(1+\sqrt{1-\left(\frac{J}{M\ell}\right)^2}\right)M\ell},\quad r_2=-\sqrt{\left(1+\sqrt{1-\left(\frac{J}{M\ell}\right)^2}\right)M\ell},\\
r_3&=\sqrt{\left(1-\sqrt{1-\left(\frac{J}{M\ell}\right)^2}\right)M\ell},\quad r_4=-\sqrt{\left(1-\sqrt{1-\left(\frac{J}{M\ell}\right)^2}\right)M\ell},
\end{align*}
where $r_1$ and $r_3$ correspond to event horizon and Cauchy horizon, i.e. physical horizons, while $r_2$ and
$r_4$ represent the negative and un-physical ``virtual'' horizons which often is discarded in literature.

The entropy at each horizon is equal to twice the perimeter length of the horizon \cite{Banados:1992wn}, i.e.
\begin{align}
S_i=4\pi r_i.
\end{align}
A straightforward calculation then gives the entropy relations. The entropy product
\begin{align}
\prod_{i=1}^4S_i=64\pi^4J^2\ell^2,
\end{align}
is strictly independent of mass $M$. It only depends on angular momentum $J$, which is consistent with the results in general rotating spacetime \cite{Cvetic:2010mn,Cvetic:2013eda}. We also present the ``part'' entropy product of BTZ black hole here
\begin{align}
\sum_{1\leq i<j\leq 4}S_iS_j=-16\pi^2M\ell^2.
\end{align}
It is mass dependent and not solely cosmological constant dependent, which is different from the case in four and high dimensional static spacetime \cite{Wang:2013nvz}. The entropy sum
\begin{align}
\sum_{i=1}^4S_i=0.
\label{entropy sum BTZ}
\end{align}
is vanishing and independent of mass $M$, and is consistent with the results in general odd dimensions \cite{Xu:2013zpa}.

However, when the $J$ is vanishing, the black hole reduces to static case with only two horizons
\begin{align*}
  r_+=\sqrt{M}\ell,\quad r_-=-\sqrt{M}\ell.
\end{align*}
In this case, entropy product and ``part'' entropy product merge into the same entropy product and behavior as
\begin{align}
  S_+S_-=-16\pi^2M\ell^2,
\end{align}
which is mass dependent. Luckily, the entropy sum Eq.(\ref{entropy sum BTZ}) still holds. Namely, there is still a entropy relation having mass independence, and revealing some features about the origin of black hole entropy at the microscopic level.

\subsection{Entropy relations of $(2+1)$ dimensional hairy black hole}
Next let us turn to the entropy relations of the static black hole with a scalar hair in $(2+1)$ dimensions constructed in \cite{Martinez:1996p2505,Nadalini:2007p2561,Xu:2013nia}. The
Lagrangian of the theory is
\begin{align}
\mathcal{L}=\frac 12\int d^3x\sqrt{-g}\left(R-g^{\mu\nu}\nabla_{\mu}\phi\,\nabla_{\nu}\phi-\xi R\phi^2-2V(\phi)\right),
\end{align}
where $\xi=\frac 18$ is the couple constant between gravity and the scalar field $\phi$. The gravitational constant $\kappa$ has been set to be $1$. The metric ansatz takes the form
\begin{align*}
ds^2=-f(r)dt^2+\frac{dr^2}{f(r)}+r^2d\psi^2,
\end{align*}
where $\psi$ is the angular coordinate ranges $-\pi\leq\psi <\pi$. However, we should not restrict the $r$
coordinate because of the ``virtual'' horizons. Here we are interested in the static uncharged
hairy AdS black hole with the horizon function \cite{Martinez:1996p2505,Nadalini:2007p2561,Xu:2013nia}
\begin{align}
f(r)&=-\frac M3 \left(3+\frac{2B}{r}\right)+\frac{r^2}{\ell^2}\label{f3mao},
\intertext{the scalar field is}
\phi(r)&=\pm\sqrt{\frac{8B}{r+B}}\nonumber,
\intertext{and self-coupling potential}
V(\phi)&=-\frac{1}{\ell^2}+\frac{1}{512}\left(\frac{1}{\ell^2}-\frac{M}{3B^2}\right)\phi^6\nonumber,
\end{align}
where $M$ is the mass of black hole and $B$ characterizes the strength of the scalar field.

Since we consider the non-minimal coupled scalar hairy black hole, the entropy of horizon does not satisfy the Benkenstein and Hawking's theorem. A non-minimal coupling factor need to be multiplied by $\frac A4$, as the entropy of horizon. Namely, the entropy for each horizon is
\begin{align}
S_i=\frac{A}{4}\left(1-\xi\phi(r_i)^2\right) \notag
   =\frac{\pi r_i}{2}\left(1-\frac{B}{r_i+B}\right)
   =\frac{\pi r_i^2}{2(r_i+B)},
\end{align}
Back to the horizon function $f(r)$, which is a cubic polynomial. It is not very difficult to solve its roots
out (See Appendix \ref{appA}). There are three horizons located at $r_1$, $r_2$ and $r_3$, what are zeros of $f(r)$, part of which may be negative, even complex. With the help of computer algebra system (CAS) one can calculate that the entropy product and ``part'' entropy product, however, are mass dependent
\begin{align}
\prod_{i=1}^3S_i&=\,\frac{\pi^3M^2\ell^4B}{6(3B^2-M\ell^2)}, \\
\sum_{1\leq i<j\leq 3}S_iS_j&=\frac{\pi^2M^2\ell^4}{4(3B^2-M\ell^2)}.
\end{align}
When $B$ is vanishing, the ``part'' entropy product reduces to the case for BTZ black hole, which is still dependent of mass. Comparing with the result in general rotating spacetime \cite{Cvetic:2010mn,Cvetic:2013eda}, one can conclude that the entropy product is always mass dependent when the charge $Q$ and $J$ is vanishing. Furthermore, the mass independence of ``part'' entropy product is not expected to hold in the case which the black holes do not admit the entropy area law. Luckily again, the entropy sum
\begin{align}
\sum_{i=1}^3S_i=0
\end{align}
is vanishing and independent of mass $M$.

As the entropy sum in general odd dimensions is vanishing and independent of mass $M$ \cite{Xu:2013zpa}, consider all theories in three dimensions, there is always a entropy relation (entropy sum) having mass independence, and revealing some features about the origin of black hole entropy at the microscopic level.

\section{Entropy relations of black holes with multi-horizons in four and higher dimensions}
In this section, we first revisit some known entropy relations and give some new unknown ones in four and higher dimensions. We show the entropy product, ``part" entropy product and entropy sum in four and higher dimensions separately. Then we discuss their differences and similarity in general dimensions, in order to make a further study on understanding the origin of black hole entropy at the microscopic level.

\subsection{Entropy product}
The entropy product of multi-horizons black hole is studied widely in many theories, including the super-gravity model \cite{Cvetic:2010mn,Toldo:2012ec,Cvetic:2013eda,Lu:2013ura,Chow:2013tia}, Einstein gravity \cite{Detournay:2012ug,Castro:2012av,Visser:2012zi,Chen:2012mh,Castro:2013kea,Visser:2012wu,Abdolrahimi:2013cza,Pradhan:2013hqa} and other modified gravity models \cite{Castro:2013pqa,Cvetic:2013eda,Faraoni:2012je,Lu:2013eoa,Anacleto:2013esa} in both four and higher dimensions. Hence, we will only summarize its features here.
\begin{enumerate}
  \item It is always independent of the mass of the black hole, and can be expressed solely in terms of the quantized charges including the electric charge $Q$, the angular momentum $J$, and the cosmological constant $\Lambda$ (which can be treated as pressure after explaining the mass of the black hole as enthalpy rather than internal energy of the system) \cite{Cvetic:2010mn,Castro:2012av,Toldo:2012ec,Chen:2012mh,Visser:2012zi,Cvetic:2013eda,Abdolrahimi:2013cza,Lu:2013ura,
Anacleto:2013esa,Chow:2013tia,Castro:2013kea,Lu:2013eoa,Wang:2013smb,Xu:2013zpa};
  \item It holds for the black holes in four and higher dimensions asymptotically flat and asymptotically (anti-)de Sitter spacetimes, including the ordinary area entropy \cite{Detournay:2012ug,Castro:2012av,Visser:2012zi,Chen:2012mh,Castro:2013kea,Visser:2012wu,Abdolrahimi:2013cza,Pradhan:2013hqa} and non-area entropy for which with higher derivative terms in the lagrangian \cite{Castro:2013pqa,Cvetic:2013eda,Faraoni:2012je,Lu:2013eoa,Anacleto:2013esa};
  \item One need to include the necessary effect of the un-physical, ``virtual'' horizons, in order to preserve its mass independence \cite{Visser:2012wu,Wang:2013smb,Xu:2013zpa}.
  \item It is shown that the charge $Q$, $J$ and $\Lambda$ plays an important role in this entropy product. When the rotating black holes reduce to static case, the mass independence of entropy product always fails \cite{Detournay:2012ug,Castro:2013pqa,Visser:2012wu}; electric charge $Q$ plays the same role with $J$ in $f(R)$-Maxwell theory, as entropy product of uncharged $f(R)$ black holes depends on mass; in asymptotically flat spacetime, the mass independence of entropy product is destroyed only in the case of the black holes in higher derivative theories, i.e. Gauss-Bonnet theory, even the charge $Q$ and $J$ are not vanishing, such as Kerr-Newman black hole in Gauss-Bonnet theory \cite{Castro:2013pqa} since a constant $\alpha$ which appears in $\mathcal{L}_{GB}$ is added in the ordinary area entropy.
\end{enumerate}

\subsection{``Part'' entropy product}
In order to preserve the mass independence of entropy relations, ``part'' entropy product are introduced \cite{Visser:2012wu,Wang:2013nvz}.  The ``part'' entropy product we have constructed is $\sum_{1\leq i<j\leq D}(S_{i}S_{j})^{\frac{1}{d-2}}$, where $D$ is the number of horizons, including physical and ``virtual'' ones; $d$ is the number of the dimensions. This type of entropy relation was firstly introduced in \cite{Visser:2012wu} in four dimensions and then generalized to general dimensions \cite{Wang:2013nvz}. However, the ``part'' entropy product could be calculated smoothly only in the Benkenstein-Hawking entropy case. Here we revisit the ``part'' entropy product in general dimensions in Table \ref{tab:part_entropy_product_relation}. One should note that, for charged black hole in $f(R)$ gravity, we only demonstrate the $d=4$ case for simplification, since the standard Maxwell energy-momentum tensor is not traceless, which makes people failed to derive higher dimensional black hole/string solutions from $f(R)$ gravity coupled to standard Maxwell field \cite{Sheykhi:2012zz}.
\begin{table}[!ht]
\makegapedcells
\begin{tabular}{cc}
\toprule
black holes & ``part'' entropy product \\
\midrule
Schwarzschild-(A)dS &
$\disy-\frac{k(d-1)(d-2)\pi^{\frac{d-1}{d-2}}}{2\Lambda\left(2\Gamma\Bigl(\frac{d-1}{2}\Bigr)\right)^{\frac{2}{d-2}}}$ \cite{Visser:2012wu,Wang:2013nvz}\\
Reissner-Nordstr{\"o}m-(A)dS & $\disy-\frac{k(d-1)(d-2)\pi^{\frac{d-1}{d-2}}}{2\Lambda\left(2\Gamma\Bigl(\frac{d-1}{2}\Bigr)\right)^{\frac{2}{d-2}}}$ \cite{Visser:2012wu,Wang:2013nvz}\\
neutral $f(R)$ & $\disy-\frac{kd(d-1)}{R_0}
     \left(\frac{\pi^{\frac{d-1}{2}}(1+f^{\prime}(R_0))}{2\Gamma\Bigl(\frac{d-1}{2}\Bigr)}\right)^{\frac{2}{d-2}}$ \cite{Wang:2013nvz}\\
charged $f(R)(d=4)$ & $\disy-\frac{12k}{R_0} (1+f^{\prime}(R_0))\pi$ \cite{Wang:2013nvz} \\
\bottomrule
\end{tabular}
\centering
\caption{``part'' entropy product of some black holes} \label{tab:part_entropy_product_relation}
\end{table}

In Table \ref{tab:part_entropy_product_relation}, $k$ is the parameter signifies the geometry of horizons and can only be $+1$, $0$ and $-1$, corresponding to spherical, flat and hyperbolic horizons. From Table \ref{tab:part_entropy_product_relation}, it can be found that, in the static spacetime, the ``part'' entropy product only depends on the cosmological constant $\Lambda$. It never depends on the conserved charges $Q$, nor even the mass $M$. In other words, the electric charge $Q$ does not play an important role (as $J$) in ``part'' entropy product, as the results have no difference between the charged and uncharged black holes. In asymptotically flat spacetime, the mass independence of ``part'' entropy product is destroyed, even the electric charge $Q$ is not vanishing, such as Einstein-Maxwell black holes in general dimensions \cite{Wang:2013nvz}. In the rotating spacetime, taking BTZ as an example (as shown in Section 2), the mass independence of the ``part'' entropy product fails. Still one need include the necessary effect of the un-physical ``virtual'' horizons, in order to preserve its mass independence.

\subsection{Entropy sum}
Entropy sum \cite{Wang:2013smb,Xu:2013zpa} is another entropy relation, which is introduced in order to preserve the mass independence of entropy relations. It is shown firstly in four dimensions \cite{Wang:2013smb} and is generalized to higher dimensions \cite{Xu:2013zpa} soon. Here we revisit entropy sum in various gravity theory and present some new unknown one in Table \ref{tab:entropy_sum_relation}. One should note that, since in $d=4$ dimensions, the integration of the GB density $\mathcal{L}_{\rm GB}=R_{\mu\nu\gamma\delta} R^{\mu\nu\gamma\delta}-4R_{\mu\nu} R^{\mu\nu}+R^2$ is a topological number and has no dynamics, which makes it out of our discussion. In the high dimensional Einstein-scalar theory and Einstein-Weyl theory, entropy sum for multi-horizons black holes are still difficult to obtain.
\begin{table}[!ht]
\makegapedcells
\begin{tabular}{ccc}
\toprule
black holes & $4$ dimensions & $6$ dimensions\\
\midrule
Schwarzschild-(A)dS & $\disy\frac{6\pi k}{\Lambda}$ \cite{Wang:2013smb} & $\disy\frac{400 k^2\pi^2}{3\Lambda^2}$ \cite{Xu:2013zpa} \\
Reissner-Nordstr{\"o}m-(A)dS & $\disy\frac{6\pi k}{\Lambda}$ \cite{Wang:2013smb} & $\disy\frac{400 k^2\pi^2}{3\Lambda^2}$ \cite{Xu:2013zpa} \\
(A)dS black hole in $f(R)$ & $\disy\frac{6\pi k}{\Lambda_f}(1+f^{\prime}(R_0))$ & $\disy\frac{400
                                                        k^2\pi^2}{3\Lambda_f^2}(1+f^{\prime}(R_0))$ \cite{Xu:2013zpa}\\
Kerr-(A)dS & $\disy\frac{6\pi}{\Lambda}$ \cite{Wang:2013smb} & $\disy\frac{4\pi^2}{3\Lambda^2}$ (Appendix \ref{appB})\\
Hairy black holes & $\disy\pm\frac{6\pi}{\Lambda}\pm\frac{\pi}{6\alpha}$ \cite{Wang:2013smb} & $--$ \\
Gauss-Bonnet (A)dS & $--$ & $\disy\frac{50k^2}{\Lambda^2}+\frac{30 k^2\tilde{\alpha}}{\Lambda}$ \cite{Xu:2013zpa}\\
Einstein-Weyl & $\begin{aligned}&4\pi\alpha(1-2c)  &\text{for charged \cite{Wang:2013smb}}  \\
                &-4\pi\alpha &\text{for neutral \cite{Wang:2013smb}}\end{aligned}$  & $--$ \\
gauged supergravity & $\disy -\frac{2\pi}{g^2}$ \cite{Xu:2013zpa}& $\disy\frac{4\pi^2}{3g^4}$ \cite{Xu:2013zpa}\\
\bottomrule
\end{tabular}
\centering
\caption{entropy sum relation for some gravity theory in four and six dimensions} \label{tab:entropy_sum_relation}
\end{table}

In Table \ref{tab:entropy_sum_relation}, $\Lambda_f$ is effective cosmological constant in $f(R)$ gravity. \noindent $k$ is the parameter signify the geometry of horizons and can only be $+1$, $0$ and $-1$,
corresponding to spherical, flat and hyperbolic horizons. We demonstrate the calculation of entropy
sum of the six dimensional Kerr-(A)dS black hole in Appendix \ref{appB}. From Table \ref{tab:entropy_sum_relation}, one can obtain that there is \textit{no} angular momentum $J$, electric charge $Q$ and mass $M$ in the entropy sum.
There only depend on the cosmological constant $\Lambda$ or $g^2$ and the constant characterizing the strength of the background spacetime, no matter in static spacetime or rotating spacetime, for charged or uncharged black holes. That is to say, angular momentum $J$ and electric charge $Q$ do not play an important role in entropy sum, as the results have no difference between the charged and uncharged, static and rotating black holes. Besides, in six dimensional AdS spactime, entropy sum of black holes admitting the area entropy law are the same with that in dS spacetime. On the other side, as shown in \cite{Xu:2013zpa}, the entropy sum in general odd dimensions is vanishing and independent of mass $M$. One still need to include the necessary effect of the un-physical ``virtual'' horizons, in order to preserve its mass independence. In asymptotically flat spacetime, the entropy sum of multi-horizons black holes is not independent of mass, even the charge $Q$ and $J$ are not vanishing, such as Kerr black hole \cite{Wang:2013smb}.

\subsection{Differences and similarities between three entropy relations}
In this subsection, we discuss the differences and similarities between entropy product, ``part'' entropy product and entropy sum in three, four and higher dimensions, in order to make a further study on explaining the origin of black hole entropy at the microscopic level.

After having a whole look at the entropy relations, one can conclude that they can be divided into two kinds barely: entropy product and ``part" entropy product belong to product relations for entropy, while entropy sum belong to sum relation. For the former relation, one can find that the mass independence of entropy product and ``part" entropy product hold complementary. In rotating (A)dS spacetime, entropy product is independent of mass (and depends on electric charge, angular momentum and cosmological constant), while reducing to static (A)dS spacetime ``part" entropy product takes the mass independence (and only depends on cosmological constant). When the case in (A)dS spacetime reduces to flat spacetime, It is back to entropy product which is independent of mass, while ``part" entropy product turns to depend on mass. Besides, they never holds in the same case. Hence they can be viewed as the same kind of universal entropy relations. Consider the mass independence, it seems that there is always one of the product relations for entropy holding. However, two kind of failed examples are known: the degenerated cases with only two horizons, whose entropy product and ``part'' entropy product merge into the same entropy product and have some vanishing charges (electric charge and angular momentum, and cosmological constant) meanwhile, such as the static uncharged BTZ black hole and Kerr-Newman black hole \cite{Castro:2013pqa}; the cases do not admit the entropy area law, such as the three dimensional static uncharged hairy black hole and Gauss-Bonnet-AdS black holes \cite{Castro:2013pqa}. For the entropy sum, it only depends on the constant characterizing the strength of the background spacetime (the cosmological constant etc.) and never depends on electric charge, angular momentum and mass. Obviously, one can immediately point out the black holes case whose entropy relations having no mass independence, namely the black hole with vanishing electric charge and angular momentum, and cosmological constant. However, for this special case, people will always find only one horizon which is not interesting, such as the Schwarzschild black holes.

On the other hand, consider the ``part'' entropy product and entropy sum together, we find that they both only depend on the constant characterizing the strength of the background spacetime (the cosmological constant etc.), other than the electric charge and angular momentum, which is different from the entropy product. It is certain that this two relations have some similarities. Actually they are equal in some sense. For the sake of brevity, we will only take the four dimensional Schwarzschild-de-Sitter black hole as an example. The calculations for the other cases (the Kerr-Newman-(Anti-)de-Sitter black holes etc.) can be performed in a similar manner. We begin with the horizon function of the four dimensional Schwarzschild-de-Sitter black hole
\begin{equation*}
f(r)=1-\frac{2M}{r}-\frac{\Lambda r^2}{3}
\end{equation*}
where $M$ is the mass of the black hole and $\Lambda$ is the cosmological constant
We will substitute $\Lambda=\frac{1}{L^2}$ for convenience here. As we are aim to the entropy relations of black hole horizons, we first list all three horizons  \cite{Visser:2012wu}
\begin{align*}
r_{E} &=2L\sin\left(\frac 13 \arcsin \left(\frac{3 M}{L}\right)\right)\\
r_{C} &=2L\sin\left(\frac 13 \arcsin \left(\frac{3 M}{L}\right)+\frac{2\pi}{3}\right)\\
r_{V} &=2L\sin\left(\frac 13 \arcsin \left(\frac{3 M}{L}\right)-\frac{2\pi}{3}\right),
\end{align*}
where, $r_{E}$ and $r_{C}$ represent the event horizon and cosmological horizon respectively, while $r_V$ is a ``virtual'' horizon. The ``part'' entropy product of Schwarzschild-de-Sitter black hole takes the following form \cite{Visser:2012wu,Wang:2013smb}
\begin{align}
  \sqrt{S_ES_C}+\sqrt{S_CS_V}+\sqrt{S_VS_E}=-3\pi L^2,
\end{align}
which can lead into the entropy relations of physical horizon having mass independence \cite{Visser:2012wu}
\begin{align}
  S_E+S_C+\sqrt{S_ES_C}=3\pi L^2.
  \label{phyrela}
\end{align}
Now we prove the entropy sum of Schwarzschild-de-Sitter black hole leads into the above relations as well. Firstly we introduce the relationship of three horizons
\begin{align*}
  r_E+r_C+r_V=0.
\end{align*}
After inserting the entropy of each horizon, i.e. $S_{i}=A_{i}/4=\pi r_i^2$, we conclude
\begin{align}
  \sqrt{S_E}+\sqrt{S_C}+\sqrt{S_V}=0,
\end{align}
which results in
\begin{align}
  S_V=S_E+S_C+2\sqrt{S_ES_C}.
  \label{ecv}
\end{align}
This makes us possible to eliminate the un-physical ``virtual'' horizons, constructing combinations of physical horizon entropy and linking the different entropy relations. Inserting the above relation into the entropy sum \cite{Wang:2013smb}
\begin{align}
  S_E+S_C+S_V=6 \pi L^2,
\end{align}
we immediately obtain the mass independent entropy relations of physical horizon Eq.(\ref{phyrela}). Hence, we conclude that ``part'' entropy product and entropy sum of Schwarzschild-de-Sitter black hole are actually equal, when only the effect of the physical horizons are considered. Back to the relation Eq.(\ref{phyrela}), it also reveal that one can explaining the origin of black hole entropy at the microscopic level without considering the effect of the un-physical virtual horizons.

We know that there was an understanding for entropy product via thermodynamics on event and Cauchy horizons, i.e. the thermodynamics correspondence shown in \cite{Chen:2012mh}. Thus the mass independence of entropy sum also has similar physical understanding because of the equality between entropy product and sum. Note the calculation of entropy sum based on three or more horizons. However, we have not yet known clearly the thermodynamics on ``virtual'' horizon(s). This makes the physical picture of entropy sum difficult to be found as done to entropy product temporarily, while we can obtain some thermodynamic properties of the ``virtual'' horizon by this procedure, including the thermodynamics laws and Smarr's relations, at least in certain solutions. In this perspective, ``entropy sum'' is of importance to black hole thermodynamics. This will be our future work.

\section{Conclusions}
In this paper, we firstly present all entropy relations include entropy product, ``part" entropy product and entropy sum in three dimensions, which is never studied in previous literature, in order to improve the study on entropy relations. Then we revisit some known entropy relations and give some new unknown ones in four and high dimensions. We discuss their differences and similarity in general dimensions, in order to make a further study on understanding the origin of black hole entropy at the microscopic level. After having a whole look at the entropy relations, we conclude
\begin{enumerate}
  \item Entropy product and ``part'' entropy product belong to the same kind of entropy relation, because the mass independence of entropy product and ``part" entropy product hold complementary in the theory without Maxwell field. In rotating (A)dS spacetime, entropy product is independent of mass, while reducing to static (A)dS spacetime ``part'' entropy product takes the mass independence. When the case in static (A)dS spacetime reduces to flat spacetime, electric charge $Q$ plays a switch role to the dependent of mass or not of entropy product, while ``part'' entropy product turns to depend on mass \cite{Wang:2013nvz}. When the Maxwell field is considered, entropy product and ``part'' entropy product may hold meanwhile, that is to say, the behaviors of entropy product is more complicated and interesting.
  \item  There are two kind of failed examples which the mass independence of ``part'' entropy product or entropy product disappears: the degenerated cases with only two horizons, i.e. the cases that entropy product and ``part'' entropy product merge into the same entropy product and have some vanishing charges (electric charge and angular momentum, and cosmological constant) meanwhile, such as the static uncharged BTZ black hole and Kerr-Newman black hole in Gauss-Bonnet theory \cite{Castro:2013pqa}; since these cases do not admit the entropy area law.
  \item ``Part" entropy product and entropy sum both have solely the constant (characterizing the strength of the background spacetime, such as the cosmological constant etc.) dependence, other than the electric charge and angular momentum dependence, which is different from the entropy product.
  \item ``Part" entropy product and entropy sum of Schwarzschild-de-Sitter black hole are actually equal, when only the effect of the physical horizons are considered, as they both can be simplified into a mass independent entropy relations of physical horizon. This also reveal that one can explaining the origin of black hole entropy at the microscopic level without considering the effect of the un-physical virtual horizons.
\end{enumerate}
However, for the sake of brevity, the idea of this paper is only checked in some simplest examples, while the calculations for the other cases can be performed in a similar manner. We hope to return to this subject in the future.

\section*{Acknowledgments}
This work is partially supported by the Natural Science Foundation of China (NSFC) under Grant No.11075078 and
by the project of knowledge innovation program of Chinese Academy of Sciences.

\appendix
\section{Solving the Cubic Polynomial}\label{appA}
This appendix presents the horizons of three dimensional static uncharged hairy black hole following the procedure introduced in \cite{Visser:2012wu}. Compare with the convenient form
\begin{align*}
r^3-3p^2r+2q=0
\end{align*}
and its three roots, accordingly
\begin{align*}
r=2p\sin\left(\frac 13\arcsin\left(\frac{q}{p^3}\right)+\epsilon\frac{2\pi}{3}\right), \quad\epsilon\in
  \{0, \pm 1\}.
\end{align*}
The horizon function Eq.(\ref{f3mao}) of three dimensional static uncharged hairy black hole could be simplified as
\begin{align*}
r^3-M\ell^2r-\frac{2M\ell^2B}{3}=0.
\end{align*}
After taking the following transformation
\begin{align*}
p=\sqrt{\frac{M\ell^2}{3}}=m\ell, \qquad q=-\frac{M\ell^2B}{3}=-m^2\ell^2B.
\end{align*}
where we have set $m^2=\frac{M}{3}$ for convenience. Then three roots can be obtained
\begin{align*}
r_1&=2m\ell\sin\left[\frac 13 \arcsin\left(-\frac{B}{m\ell}\right)\right], \\
r_2&=2m\ell\sin\left[\frac 13 \arcsin\left(-\frac{B}{m\ell}\right)+\frac{2\pi}{3}\right], \\
r_3&=2m\ell\sin\left[\frac 13 \arcsin\left(-\frac{B}{m\ell}\right)-\frac{2\pi}{3}\right].
\end{align*}
Note that not all of them are physical. However, we can choose the parameters $M$ and $B$ to guarantee one
(event) or two (event and Cauchy) positive horizon(s), while the rest are(is) negative and un-physical. It is shown that the un-physical ``virtual'' horizons is necessary, in order to preserve mass independence of the entropy relations.

\section{Calculation of the Entropy Sum of the $6$ Dimensional Kerr-(A)dS Black Hole}\label{appB}
Arbitrary dimensions Kerr-(A)dS black hole was constructed in \cite{Gibbons:2004uw}. Here we demonstrate the
calculation of entropy sum for the $6$ dimensional case with the horizon function
\begin{align}
f(r)=(r^2+a^2)(r^2+b^2)(1\mp\Lambda r^2)-2Mr,
\end{align}
where the upper and lower of sign stand for the dS and AdS solution respectively. And $a$ and $b$ are independent rotation parameters. Set $f(r)=0$ and reform it to polynomial
\begin{align} \label{eq:horizon_function_6_D_KAdS}
\mp\Lambda r^6+\bigg(\mp\Lambda(a^2+b^2)+1\bigg)r^4+\bigg(\mp\Lambda a^2b^2+a^2+b^2\bigg)r^2-2Mr+a^2b^2=0.
\end{align}
In principle, the polynomial has six zeros corresponding to six horizons, including un-physical ``virtual'' horizons. By using the Vieta theorem on (\ref{eq:horizon_function_6_D_KAdS}), it is easy to obtain some useful relationships
\begin{align} \label{eqs:roots_sum}
&\sum_{i=1}^6 r_i=0, \qquad \sum_{1\leq i<j\leq6}r_ir_j=\frac{\mp\Lambda(a^2+b^2)+1}{\mp\Lambda}, \notag \\
&\sum_{1\leq i<j<k<l\leq6}r_ir_jr_kr_l=\frac{\mp\Lambda a^2b^2+a^2+b^2}{\mp\Lambda}.
\end{align}
Consider the entropy for each horizon
\begin{align}
S_i&=\frac{A_i}{4}=\frac{1}{4}\frac{2\pi^{\frac 52}(r_i^2+a^2)(r_i^2+b^2)}{\Gamma\left(\frac 52\right)(1\pm \Lambda a^2)(1\pm \Lambda b^2)}\nonumber\\
   &=\frac{2\pi^2}{3(1\pm \Lambda a^2)(1\pm \Lambda b^2)}\bigg(r_i^4+(a^2+b^2)r_i^2+a^2b^2\bigg).
\end{align}
In order to obtain the entropy sum of all horizons, we firstly calculate $\disy\sum_{i=1}^6 r_i^4$ and $\disy\sum_{i=1}^6 r_i^2$ separately with the help of the following universal equalities
\begin{align}
3\sum_{i=1}^6 r_i^4=&4\left(\sum_{i=1}^6 r_i\right)\left(\sum_{i=1}^6 r_i^3\right)
                    +6\left(\sum_{1\leq i<j\leq6}r_ir_j\right)^2 \notag \\
                    &-\left(\sum_{i=1}^6 r_i\right)^4
                     -12\left(\sum_{1\leq i<j<k<l\leq6}r_ir_jr_kr_l\right) \nonumber\\
\sum_{i=1}^6 r_i^2=&\left(\sum_{i=1}^6 r_i\right)^2-2\left(\sum_{1\leq i<j\leq6}r_ir_j\right). \nonumber
\end{align}
Inserting Eq.(\ref{eqs:roots_sum}) into the above two equalities, straightforward we get the entropy sum of six dimensions Kerr-(A)dS black hole
\begin{align}
\sum_{i=1}^6S_i=\frac{4\pi^2}{3\Lambda^2}.
\end{align}

\providecommand{\href}[2]{#2}\begingroup
\small\itemsep=0pt
\providecommand{\eprint}[2][]{\href{http://arxiv.org/abs/#2}{arXiv:#2}}


\begin{thebibliography}{99}


\bibitem{Cvetic:2010mn}
  M.~Cvetic, G.~W.~Gibbons and C.~N.~Pope,
  ``Universal Area Product Formulae for Rotating and Charged Black Holes in Four and Higher Dimensions,''
  Phys.\ Rev.\ Lett.\  {\bf 106}, 121301 (2011)
  [\eprint{1011.0008}].

\bibitem{Toldo:2012ec}
  C.~Toldo and S.~Vandoren,
  ``Static nonextremal AdS4 black hole solutions,''
  JHEP {\bf 1209}, 048 (2012)
  [\eprint{1207.3014}].

\bibitem{Cvetic:2013eda}
  M.~Cvetic, H.~Lu and C.~N.~Pope,
  ``Entropy-Product Rules for Charged Rotating Black Holes,''
  Phys.\ Rev.\ D {\bf 88}, 044046 (2013)
  [\eprint{1306.4522}].

\bibitem{Lu:2013ura}
  H.~Lu, Y.~Pang and C.~N.~Pope,
  ``AdS Dyonic Black Hole and its Thermodynamics,''
  [\eprint{1307.6243}].

\bibitem{Chow:2013tia}
  D.~D.~K.~Chow and G.~Comp¨¨re,
  ``Seed for general rotating non-extremal black holes of N=8 supergravity,''
  [\eprint{1310.1925}].

\bibitem{Detournay:2012ug}
  S.~Detournay,
  ``Inner Mechanics of 3d Black Holes,''
  Phys.\ Rev.\ Lett.\  {\bf 109}, 031101 (2012)
  [\eprint{1204.6088}].

\bibitem{Castro:2012av}
  A.~Castro and M.~J.~Rodriguez,
  ``Universal properties and the first law of black hole inner mechanics,''
  Phys.\ Rev.\ D {\bf 86}, 024008 (2012)
  [\eprint{1204.1284}].

\bibitem{Visser:2012zi}
  M.~Visser,
  ``Quantization of area for event and Cauchy horizons of the Kerr-Newman black hole,''
  JHEP {\bf 1206}, 023 (2012)
  [\eprint{1204.3138}].

\bibitem{Chen:2012mh}
  B.~Chen, S.~-x.~Liu and J.~-j.~Zhang,
  ``Thermodynamics of Black Hole Horizons and Kerr/CFT Correspondence,''
  JHEP {\bf 1211}, 017 (2012)
  [\eprint{1206.2015}].

\bibitem{Castro:2013kea}
  A.~Castro, J.~M.~Lapan, A.~Maloney and M.~J.~Rodriguez,
  ``Black Hole Monodromy and Conformal Field Theory,''
  Phys.\ Rev.\ D {\bf 88}, 044003 (2013)
  [\eprint{1303.0759}].

\bibitem{Visser:2012wu}
  M.~Visser,
  ``Area products for black hole horizons,''
  Phys.\ Rev.\ D {\bf 88}, 044014 (2013)
  [\eprint{1205.6814}].

\bibitem{Abdolrahimi:2013cza}
  S.~Abdolrahimi and A.~A.~Shoom,
  ``Distorted Five-dimensional Electrically Charged Black Holes,''
  [\eprint{1307.4406}].

\bibitem{Pradhan:2013hqa}
  P.~Pradhan,
  ``Area Products and Mass Formula for Kerr Newman Taub Nut Spacetime,''
  [\eprint{1310.7921}].  

\bibitem{Castro:2013pqa}
  A.~Castro, N.~Dehmami, G.~Giribet and D.~Kastor,
  ``On the Universality of Inner Black Hole Mechanics and Higher Curvature Gravity,''
  [\eprint{1304.1696}].

\bibitem{Faraoni:2012je}
  V.~Faraoni and A.~F.~Z.~Moreno,
  ``Are quantization rules for horizon areas universal?,''
  [\eprint{1208.3814}].

\bibitem{Lu:2013eoa}
  H.~Lu,
  ``Charged dilatonic ads black holes and magnetic AdS$_{D-2} \times R^{2}$ vacua,''
  JHEP {\bf 1309}, 112 (2013)
  [\eprint{1306.2386}].

\bibitem{Anacleto:2013esa}
  M.~A.~Anacleto, F.~A.~Brito and E.~Passos,
  ``Acoustic Black Holes and Universal Aspects of Area Products,''
  [\eprint{1309.1486}].

\bibitem{Wang:2013nvz}
  J.~Wang, W.~Xu and X.~-h.~Meng,
  ``The Entropy Relations of Black Holes with Multihorizons in Higher Dimensions,''
  Phys.\ Rev.\ D {\bf 89}, 044034 (2014)
  [\eprint{1312.3057}].  

\bibitem{Wang:2013smb}
  J.~Wang, W.~Xu and X.~-h.~Meng,
  ``The ``universal property'' of Horizon Entropy Sum of Black Holes in Four Dimensional Asymptotical (anti-)de-Sitter Spacetime Background,''
  JHEP {\bf 1401}, 031 (2014)
   [\eprint{1310.6811}].  

\bibitem{Xu:2013zpa}
  W.~Xu, J.~Wang and X.~-h.~Meng,
  ``\,``Entropy sum'' of (A)dS Black Holes in Four and Higher Dimensions,''
  [\eprint{1310.7690}].

\bibitem{Cvetic:1997uw}
  M.~Cvetic and F.~Larsen,
  ``General rotating black holes in string theory: Grey body factors and event horizons,''
    Phys.\ Rev.\ D {\bf 56} (1997) 4994
    [\eprint{hep-th/9705192}].  

\bibitem{Cvetic:1997xv}
  M.~Cvetic and F.~Larsen,
  ``Grey body factors for rotating black holes in four-dimensions,''
    Nucl.\ Phys.\ B {\bf 506} (1997) 107
    [\eprint{hep-th/9706071}].  

\bibitem{Cvetic:2009jn}
  M.~Cvetic and F.~Larsen,
  ``Greybody Factors and Charges in Kerr/CFT,''
  JHEP {\bf 0909} (2009) 088
  [\eprint{0908.1136}].  

\bibitem{Ansorg:2010ru}
  M.~Ansorg, J.~Hennig and C.~Cederbaum,
  ``Universal properties of distorted Kerr-Newman black holes,''
  Gen.\ Rel.\ Grav.\  {\bf 43} (2011) 1205
  [\eprint{1005.3128}].  

\bibitem{Hennig:2009aa}
  J.~Hennig and M.~Ansorg,
  ``The Inner Cauchy horizon of axisymmetric and stationary black holes with surrounding matter in Einstein-Maxwell theory: Study in terms of soliton methods,''
  Annales Henri Poincare {\bf 10} (2009) 1075
  [\eprint{0904.2071}].  

\bibitem{Ansorg:2009yi}
  M.~Ansorg and J.~Hennig,
  ``The Inner Cauchy horizon of axisymmetric and stationary black holes with surrounding matter in Einstein-Maxwell theory,''
  Phys.\ Rev.\ Lett.\  {\bf 102} (2009) 221102
  [\eprint{0903.5405}].  

\bibitem{Hennig:2008zy}
  J.~Hennig, C.~Cederbaum and M.~Ansorg,
  ``A Universal inequality for axisymmetric and stationary black holes with surrounding matter in the Einstein-Maxwell theory,''
  Commun.\ Math.\ Phys.\  {\bf 293} (2010) 449
  [\eprint{0812.2811}].  

\bibitem{Ansorg:2008bv}
  M.~Ansorg and J.~Hennig,
  ``The Inner Cauchy horizon of axisymmetric and stationary black holes with surrounding matter,''
  Class.\ Quant.\ Grav.\  {\bf 25} (2008) 222001
  [\eprint{0810.3998}].  

\bibitem{Ansorg:2007fh}
  M.~Ansorg and H.~Pfister,
  ``A Universal constraint between charge and rotation rate for degenerate black holes surrounded by matter,''
  Class.\ Quant.\ Grav.\  {\bf 25} (2008) 035009
  [\eprint{0708.4196}].  

\bibitem{Jaramillo:2007mi}
  J.~L.~Jaramillo, N.~Vasset and M.~Ansorg,
  ``A Numerical study of Penrose-like inequalities in a family of axially symmetric initial data,''
  [\eprint{0712.1741}].  

\bibitem{Booth:2007wu}
  I.~Booth and S.~Fairhurst,
  ``Extremality conditions for isolated and dynamical horizons,''
  Phys.\ Rev.\ D {\bf 77} (2008) 084005
  [\eprint{0708.2209}].  

\bibitem{Hennig:2010hz}
  J.~Hennig and G.~Neugebauer,
  ``Non-existence of stationary two-black-hole configurations,''
  [\eprint{1002.1818}].  

\bibitem{Pradhan:2013xha}
  P.~Pradhan,
  ``Black Hole Interior Mass Formula,''
  [\eprint{1310.7126}].  

\bibitem{Banados:1992wn}
  M.~Banados, C.~Teitelboim and J.~Zanelli,
  ``The Black hole in three-dimensional space-time,''
  Phys.\ Rev.\ Lett.\  {\bf 69}, 1849 (1992)
  [\eprint{hep-th/9204099}].

\bibitem{Martinez:1996p2505}
  C.~Martinez and J.~Zanelli,
  ``Conformally dressed black hole in (2+1)-dimensions,''
  Phys.\ Rev.\ D {\bf 54}, 3830 (1996)  [\eprint{gr-qc/9604021}].

\bibitem{Henneaux:2002wm}
  M.~Henneaux, C.~Martinez, R.~Troncoso and J.~Zanelli,
  ``Black holes and asymptotics of 2+1 gravity coupled to a scalar field,''
  Phys.\ Rev.\ D {\bf 65}, 104007 (2002)
  [\eprint{hep-th/0201170}].

\bibitem{Nadalini:2007p2561}
  M.~Nadalini, L.~Vanzo, and S.~Zerbini,
  ``Thermodynamical properties of hairy
  black holes in n spacetimes dimensions,''
  Phys.Rev.D77:024047,2008 [\eprint{0710.2474}].

\bibitem{Xu:2013nia}
  W.~Xu and L.~Zhao,
  ``Charged black hole with a scalar hair in (2+1) dimensions,''
  Phys.\ Rev.\ D {\bf 87}, 124008 (2013)
  [\eprint{1305.5446}].

\bibitem{Ida}
  D.~Ida,
  ``No black hole theorem in three-dimensional gravity,''
  Phys. Rev. Lett. 85 (2000) 3758 [\eprint{gr-qc/0005129}].

\bibitem{Sheykhi:2012zz}
  A.~Sheykhi,
  ``Higher-dimensional charged $f(R)$ black holes,''
  Phys.\ Rev.\ D {\bf 86}, 024013 (2012)
  [\eprint{1209.2960}].

\bibitem{Gibbons:2004uw}
  G.~W.~Gibbons, H.~Lu, D.~N.~Page and C.~N.~Pope,
  ``The General Kerr-de Sitter metrics in all dimensions,''
  J.\ Geom.\ Phys.\  {\bf 53}, 49 (2005)
  [\eprint{hep-th/0404008}].

\end{thebibliography}
\end{document}